\documentclass[conference, a4paper, twocolumn]{IEEEtran}
\usepackage{cite}
\usepackage{amsmath,amssymb,amsfonts}
\usepackage{gensymb}
\usepackage{algorithmic}
\usepackage{graphicx}
\usepackage{textcomp}
\usepackage{xcolor}

\def\BibTeX{{\rm B\kern-.05em{\sc i\kern-.025em b}\kern-.08em
    T\kern-.1667em\lower.7ex\hbox{E}\kern-.125emX}}

\DeclareMathOperator{\sinc}{sinc}
\DeclareMathOperator{\rect}{rect}

\begin{document}

\title{Beamforming with Oversampled Time-Modulated Arrays}

\author{
    \IEEEauthorblockN{Marcin Wachowiak\IEEEauthorrefmark{1}\IEEEauthorrefmark{2}, 
    André Bourdoux\IEEEauthorrefmark{1}, 
    Sofie Pollin\IEEEauthorrefmark{2}\IEEEauthorrefmark{1}}
    \IEEEauthorblockN{
        \IEEEauthorrefmark{1} imec, Kapeldreef 75, 
        3001 Leuven, Belgium \\
        \IEEEauthorrefmark{2} Department of Electrical Engineering, 
        KU Leuven, Belgium \\
        Email: marcin.wachowiak@imec.be 
        }   
}

\maketitle

\begin{abstract}
The time-modulated array (TMA) is a simple array architecture in which each antenna is connected via a multi-throw switch. The switch acts as a modulator switching state faster than the symbol rate. The phase shifting and beamforming is achieved by a cyclic shift of the periodical modulating signal across antennas. In this paper, the TMA mode of operation is proposed to improve the resolution of a conventional phase shifter. The TMAs are analyzed under constrained switching frequency being a small multiple of the symbol rate. The presented generic signal model gives insight into the magnitude, phase and spacing of the harmonic components generated by the quantized modulating sequence. It is shown that the effective phase-shifting resolution can be improved multiplicatively by the oversampling factor ($O$) at the cost of introducing harmonics. Finally, the array tapering with an oversampled modulating signal is proposed. The oversampling provides $O+1$ uniformly distributed tapering amplitudes.
\end{abstract}


\section{Introduction}

\subsection{Problem Statement}

Antenna arrays are becoming a key component of the current and future wireless networks. The high array gain combined with agile beamforming allows to compensate for the high attenuation in millimeter-wave and sub-terahertz bands, opening doors to greater bandwidth \cite{6g_vision}. However, the growing number of antenna elements entails a proportional increase in the complexity and number of radio front-end chips. To facilitate the widespread adoption of antenna arrays, more cost- and energy-efficient solutions are needed \cite{unconvent_array_arch}.

The time-modulated array (TMA) is an array architecture in which each antenna is preceded by a multi-throw RF switch that acts as a low-resolution phase shifter. The switch operates at a frequency equal or higher than the symbol rate of the transmitted signal. The simple hardware architecture of the aforementioned building block offers savings in terms of power, cost and size of the radio front-end \cite{rfic_book}, \cite{mmw_tech_book} and allows to leverage the upscaling of the antenna array systems. By adjusting the modulating sequence for each antenna, the TMAs can synthesize and steer the beam pattern \cite{overview_tma_fda}, \cite{tma_survey}.

To facilitate more widespread use of the TMA architecture, it has to be considered for wideband signals. For large bandwidths, the switching frequency becomes constrained by the hardware giving rise to previously not studied limitations. The constrained switching frequency introduces a trade-off between the spacing of the harmonic components and the improvement in the phase-shifting resolution. The trade-offs and gains need to be addressed in detail to fully quantify the potential of the TMA in practical applications with wideband signals. 

\begin{figure}[t]
    \centering
    \includegraphics[width=3.5in]{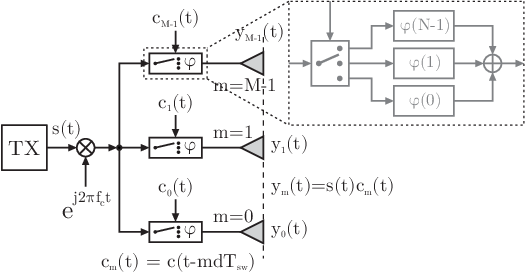}
    \caption{Architecture of the time-modulated array. Each antenna is preceded by an $N$-state switch which acts as a discrete phase shifter. The delay of the modulating sequence, which creates the modulating signal, is adjusted per antenna facilitating digital beamforming.}
    \label{fig:tma_diagram}
\end{figure}

\subsection{Relevant works}

The idea of using time modulation for beamforming originated in \cite{tma_beamscanning_init}. In \cite{tma_seq_sr_optimization}, the modulating sequences were optimized to reduce the sidelobes of a fixed beam.
Next, to allow beamforming at the center frequency and improve efficiency, the phase modulated array (PMA) was introduced in \cite{phase_tma_init}. The on-off switches were replaced by switched delay lines. 
In \cite{single_sideband_phase_iq_tma}, a TMA architecture with switched IQ modulator per antenna is investigated providing refined control over the phase shift. The analytical formulation for minimizing the undesired harmonics was derived and applied.
Simultaneous amplitude and phase weighting is studied in \cite{tma_ampl_phase_weighting_beamfomring} where it is used to reduce the sideband radiation and control the sidelobes of the beam pattern.
In \cite{lin_phase_approx_tma_eff_analysis} the phase modulation of TMA is investigated in detail with theoretical derivations of efficiency depending on the number of phase shift values. The peak-to-harmonic power ratio and sidelobe levels are studied as a function of the number of quantization steps.
In \cite{high_res_phase_shifting_w_tma}, a TMA constructed of a cascade of two 1-bit phase shifters is considered, which are controlled by two asynchronous signals. Instead of changing the pulse width of the two signals, the delay between them is adjusted. It results in improved phase-shifting resolution at the output of the modulator due to the limited timing resolution of the hardware.

Most of the listed papers evaluated the TMA in a narrowband scenario when the switching frequency is much higher than the signal bandwidth. As a consequence, the timing resolution of the control sequence and switching instants are considered arbitrary resulting in ideal and continuous control over the phase shift. However, when envisioning the TMAs for wideband signals, the quantized timing resolution needs to be taken into account.

\subsection{Contributions}

This paper evaluates the beamforming with TMA with constrained switching frequency being a small multiple of the symbol rate. The two competing uses of oversampling are studied; one increases the spacing of the harmonic components and the other improves the effective phase-shifting resolution. Finally, amplitude and beam tapering with the oversampled signal is proposed and studied.

\section{Signal processing model}

\subsection{Time-modulated array}

Consider a uniform linear array (ULA) with $M$ antenna elements. The index of the antennas is $m\in \{0, 1, \ldots, M-1\}$.
Each antenna is connected to a shared radio front-end with a dedicated $N$ throw/state switch. The hardware architecture of the TMA is illustrated in Fig. \ref{fig:tma_diagram}.
Each output of the switch is connected to a fixed phase shifter (or delay line) so that each state of the switch corresponds to a phase shift that is uniformly distributed on the unit circle.
The phase shift of $n$-th state is given by
\begin{equation}
    \label{eq:phase_shift_per_idx}
    \varphi(n) = 2 \pi \frac{n}{N},
\end{equation}
where $n$ is the index of the state $n \in \{0, 1, \ldots, N-1 \}$.
The baseband signal transmitted by the shared radio front end is denoted by $s(t)$ with bandwidth $B$ and sample rate $f_{\mathrm{s}} = B$.
To achieve time modulation, the antenna switches operate at frequency $f_{\mathrm{sw}} = O f_{\mathrm{s}}$, $O \in \mathbb{N}^+$, which is equal or greater than the sampling frequency $f_{\mathrm{s}}$ \cite{tma_pwr_freq_analysis}. 
The modulating signal is a periodically repeated sequence of $N$ rectangular pulses with a phase given by \eqref{eq:phase_shift_per_idx} which results in maximization of the power at the first harmonic \cite{lin_phase_approx_tma_eff_analysis}.

The oversampling of the modulating signal in TMAs can be utilized in two ways. It can be used either to reduce the pulse duration and increase the pulse frequency and improve the spacing of the harmonics. Or to extend the pulse duration by repeating the state to obtain finer timing and delay resolution, allowing control of the delay of the modulating sequence more accurately.
Considering the two options the oversampling factor $O$ can be factorized into 
\begin{equation}
    \label{eq:sw_freq}
    O = O_{\mathrm{f}} O_{\mathrm{\tau}} \quad O, O_{\mathrm{f}}, O_{\mathrm{\tau}} \in \mathbb{N}^+,
\end{equation}
where $O_{\mathrm{f}}$ is the modulating frequency scaling factor and $O_{\mathrm{\tau}}$ is the pulse duration scaling factor.
The pulse duration of the modulating signal is defined as
\begin{equation}
    \label{eq:pulse_duration}
    T_{\mathrm{p}} = O_{\mathrm{\tau}} \frac{1}{f_{\mathrm{sw}}} = \frac{O_{\mathrm{\tau}}}{O f_{\mathrm{s}}}.
\end{equation}
The pulse frequency is defined as
\begin{equation}
    \label{eq:mod_freq}
    f_{\mathrm{p}} = O_{\mathrm{f}} f_{\mathrm{s}}.
\end{equation}

\begin{figure}[b]
    \centering
    \includegraphics[width=3.5in]{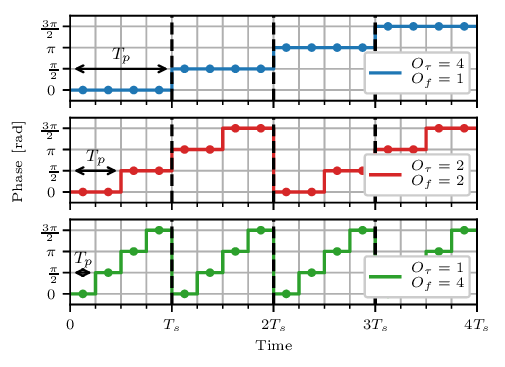}
    \caption{Modulating signal in the time domain with upsampling factor $O=4$ and number of phase shift values $N=4$ for selected pulse duration and modulation frequency scaling factors, where $T_\mathrm{s} = 1/f_{\mathrm{s}}$ is the sample period.}
    \label{fig:oversampl_sw_sig_in_td}
\end{figure}

\subsection{Modulating signal}
Due to the discretized time and phase shift values, the basic building block of the TMA modulating signal is a rectangular pulse. Consider a periodic rectangular pulse of duration $T_{\mathrm{p}}$ with unit amplitude and phase shift of $\varphi(n)$. The time domain representation of the pulse is given by 
\begin{equation}
    \label{eq:td_rect_pulse}
    g_n(t) = e^{j\varphi(n)} \sum_{i=-\infty}^{\infty} \rect{\left( \frac{t}{T_{\mathrm{p}}} - \frac{T_{\mathrm{p}}}{2} + i NT_{\mathrm{p}} \right)}.
\end{equation}
The pulse is periodic with a period $NT_{\mathrm{p}}$ which allows to expand it into the Fourier series
\begin{equation}
    \label{eq:sig_fourier_rep}
    g_n(t) = \sum_{k=-\infty}^{\infty} G_n(k) e^{j2\pi k f_{\mathrm{p}} t}.
\end{equation}
The Fourier coefficient of the $k$-th harmonic component is given by
\begin{align}
    \label{eq:fourier_coeffs}
    G_n(k) &= \frac{1}{T_{\mathrm{p}}} \int_{0}^{T_{\mathrm{p}}} g_n(t) e^{-j2\pi k f_{\mathrm{p}} t} \,dt \nonumber \\
    &= e^{j\varphi(n)} \sinc{\left(\pi \frac{k}{N} \right)} e^{-j\pi \frac{k}{N}}.
\end{align}
The modulating (control) signal $c(t)$ is composed of a sequence of $N$ delayed complex pulses from \eqref{eq:td_rect_pulse} resulting in
\begin{align}
    \label{eq:pulse_seq_td}
    c(t) &= \sum_{n=0}^{N-1} g_n \left(t - n T_{\mathrm{p}}\right).
\end{align}
Fig. \ref{fig:oversampl_sw_sig_in_td} presents the modulating signal in the time domain for selected pulse duration and modulating frequency oversampling factors. The factor $O_{\mathrm{\tau}}$ stretches the modulating signal in the time domain by replicating each state after itself $O_{\mathrm{\tau}}$ times, allowing for a finer temporal resolution.
The frequency domain representation of the modulating signal is then
\begin{align}
    \label{eq:fourier_coeffs_pulse_seq}
    C(k) &= \frac{1}{N} \sum_{n=0}^{N-1} G_n(k) e^{-j2\pi \frac{n}{N} T_{\mathrm{p}} f_{\mathrm{p}} } \nonumber \\
    &= \sinc{\left(\pi \frac{k}{N} \right)} e^{-j\pi \frac{k}{N}} \frac{1}{N} \sum_{n=0}^{N-1} e^{j2\pi \frac{n}{N} \left( 1 - k \right)} \nonumber \\
    &= \sinc{\left(\pi \frac{k}{N} \right)} e^{-j\pi \frac{k}{N}} I(k)
\end{align}
$I(k) = \frac{1}{N} \sum_{n=0}^{N-1} e^{j2\pi \frac{n}{N} \left( 1 - k \right)}$ determines the existence of the $k$-th spectral component at frequency $f_k = \frac{k}{N} f_{\mathrm{p}}$
\begin{align}
    \label{eq:fourier_comp_existence_discr_freq}
    I(k) &= \begin{cases} 
        1, & k = 1 + iN \\ 
        0, & k \neq 1 + iN
    \end{cases},
    \quad i \in \mathbb{Z}
\end{align}

\begin{figure}[b]
    \centering
    \includegraphics[width=3.5in]{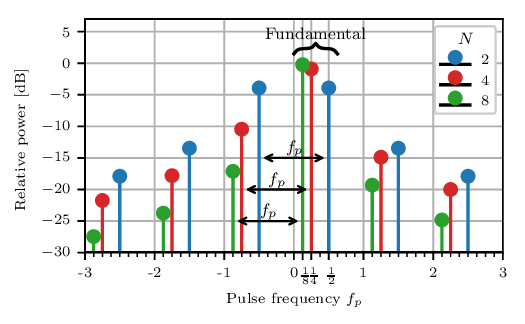}
    \caption{Power per harmonic component for modulating sequences with a selected number of phase shift values $N = 2, 4, 8$.}
    \label{fig:h_pwr_freq_per_n_phase_vals}
\end{figure}

To simplify the following analysis, the discrete harmonic representation of the modulating sequence is transformed into a continuous frequency range as follows
\begin{align}
    \label{eq:mod_seq_fd_continious}
     C(f) &= \sinc{\left(\pi \frac{f}{f_{\mathrm{p}}} \right)} e^{-j\pi \frac{f}{f_{\mathrm{p}}}} \delta \left( f - \left(\frac{f_{\mathrm{p}}}{N} + i f_{\mathrm{p}} \right)\right) \nonumber \\
    &= \alpha(i) \delta \left( f  - \frac{f_{\mathrm{p}}}{N} - i f_{\mathrm{p}} \right)
\end{align}
where $\alpha(i) = \sinc{\left(\pi \left( i + \frac{1}{N} \right) \right)} e^{-j\pi \left( i + \frac{1}{N}  \right) }$ is the complex coefficient determining the amplitude and phase of the $i$-th harmonic component and $\delta$ is the Dirac delta function.
The harmonic components of the signal are spaced in frequency by $f_{\mathrm{p}} = O_{\mathrm{f}} f_{\mathrm{s}}$, which is the pulse frequency.
The length of the sequence $N$ affects the power of the harmonics and frequency shift of the total signal, equal to modulating frequency $f_{\mathrm{mod}} = f_{\mathrm{p}} / N$. Fig. \ref{fig:h_pwr_freq_per_n_phase_vals} illustrates the power of the harmonic components in the frequency domain for a selected number of phase shift values.

\begin{figure}[b]
    \centering
    \includegraphics[width=3.5in]{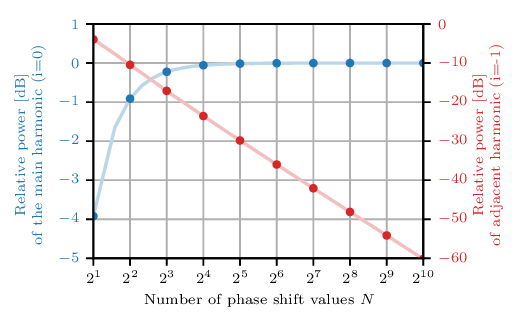}
    \caption{Power of the main harmonic $(i=0)$ and strongest undesired harmonic $(i=-1)$ as a function of the number of phase shift values.}
    \label{fig:h_pwr_vs_nphase}
\end{figure}

Based on \eqref{eq:mod_seq_fd_continious} the power of the $i$-th harmonic component located at $\frac{f_{\mathrm{p}}}{N} + i f_{\mathrm{p}}$ can be calculated as
\begin{equation}
    \label{eq:harm_pwr}
    \left| \alpha(i) \right|^2 = \left| \sinc{ \left( \pi \left( \frac{1}{N} + i \right) \right) } \right| ^2,\ i  \in \mathbb{Z}.
\end{equation}
It can be inferred from the formula that the power of the zeroth (main) harmonic component increases with the number of phase shift values $N$, while at the same time the power of the higher-order harmonics ($i \neq 0$) decreases. By increasing $N$, the power of the main harmonic and so beamforming efficiency is improved alongside the suppression of undesired harmonic components.
Fig. \ref{fig:h_pwr_vs_nphase} shows the power of the main harmonic $(i=0)$ and the highest power adjacent harmonic $(i=-1)$ as a function $N$. The power loss of the main harmonic becomes negligible for $N>=8$.

\subsection{Phase shifting with modulating signal}
The phase shift of the signal transmitted by the TMA antenna is achieved by a cyclic shift of the modulating sequence. Given a sequence of length $N$ and pulse duration $T_{\mathrm{p}}$ there are $D$ possible discrete delay values given by
\begin{equation}
    \label{eq:num_cyclic_shift_states}
    D = N \frac{T_{\mathrm{p}}}{T_{\mathrm{sw}}} = N O_{\mathrm{\tau}}
\end{equation}
The cyclic shift of the time domain modulating sequence is equivalent to the phase rotation in the frequency domain as follows
\begin{align}
    \label{eq:time_delay_mod_sig_short}
    C(f, d) &= \mathcal{F}\left\{ c \left( t - d T_{\mathrm{sw}} \right) \right\} \nonumber \\
    &= C(f) e^{-j2\pi d T_{\mathrm{sw}} f},
\end{align}
where $d \in \{0,1, \ldots, D-1\}$ is the index of selected digital delay.
By substituting \eqref{eq:mod_seq_fd_continious} in \eqref{eq:time_delay_mod_sig_short} the full form of the shifted modulating signal is obtained 
\begin{align}
\label{eq:time_delay_mod_sig_long}
    C(f, d) &= \alpha(i) e^{-j2\pi d T_{\mathrm{sw}} f} \delta \left( f - \frac{f_{\mathrm{p}}}{N} - i f_{\mathrm{p}} \right) \nonumber \\
    &= \alpha(i) e^{j\phi(i, d)} \delta \left( f - \frac{f_{\mathrm{p}}}{N} - i f_{\mathrm{p}} \right),
\end{align}
where
\begin{equation}
  \label{eq:ph_shift_per_harm_per_delay} 
  \phi(i, d) = -2\pi \frac{d}{D} \left(1 + Ni \right)  
\end{equation}
is the phase shift of the $i$-th harmonic component due to digital delay $d$.
Next, a shared front-end signal $s(t)$ is modulated by the modulating  signal resulting in the transmitted signal $y_d (t)$
\begin{equation}
    \label{eq:modulation_td}
    y_d(t) = s(t) c{\left( t - d T_{\mathrm{sw}} \right)}.
\end{equation}
As the baseband signal is bandlimited to $B$ its frequency representation can be formulated as
\begin{equation}
    \label{eq:tx_sig_fd}
    S(f) = \rect{ \left( \frac{f}{f_{\mathrm{s}}} \right) }.
\end{equation}
The modulation in the time domain is equivalent to convolution in the frequency domain
\begin{align}
    \label{eq:modulation_fd_conv}
    Y(f, d) &= S(f) * C(f, d),
\end{align}
where $*$ is the convolution operator.
By calculating the convolution the transmitted signal simplifies to
\begin{align}
    \label{eq:modulation_fd_full_conv}
    Y(f, d) &= \int_{-\infty}^{\infty} S(f - v) C(v, d) \, dv \nonumber \\ 
    &= \alpha(i) e^{j \phi(i, d)} \int_{-\infty}^{\infty} S(f -v)\delta{\left( v - \frac{f_{\mathrm{p}}}{N} - i f_{\mathrm{p}} \right)}  \, dv \nonumber \\
    &= \alpha(i) e^{j \phi(i, d)} S\left( f -\frac{f_{\mathrm{p}}}{N} - i f_{\mathrm{p}}\right).
\end{align}
The modulation shifts the baseband signal in frequency by $\frac{f_{\mathrm{p}}}{N}$ and introduces harmonic replicas with different phases and amplitudes at intervals $f_{\mathrm{p}}$. The phase of the transmitted signal is directly controlled by adjusting the digital delay per antenna.

\begin{figure}[b]
    \centering
    \includegraphics[width=3.5in]{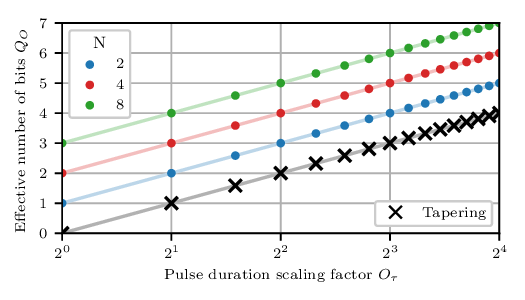}
    \caption{Effective number of bits as a function of pulse duration scaling factor $O_{\mathrm{\tau}}$ for a selected number of phase shifts $N$.}
    \label{fig:n_bits_vs_oversamp}
\end{figure}

For a system without oversampling, the number of possible phase shifts is equal to the number of phase states, similarly to that of the system without switching. However, when oversampling is considered, the pulse duration extension factor $O_{\mathrm{\tau}}$ multiplies the number of achievable phase values.
The effective phase shifter resolution is
\begin{equation}
    \label{eq:ph_shift_res}
    \Delta_{\varphi} = \frac{2\pi}{D} = \frac{2\pi}{N O_{\mathrm{\tau}}}.
\end{equation}
Increasing the switching frequency of the discrete phase shifter allows us to improve the phase-shifter resolution. The number of bits in the oversampled phase shifter is $Q_{\text{O bits}} = Q + \log_2 {O_{\mathrm{\tau}}}$, where $Q$ is the number of bits of the conventional switched phase shifter. Fig. \ref{fig:n_bits_vs_oversamp} presents the gain in the effective number of bits as a function of the oversampling factor $O_{\mathrm{\tau}}$. 

\subsection{Beamforming with TMA}

\begin{figure}[b]
    \centering
    \includegraphics[width=3.5in]{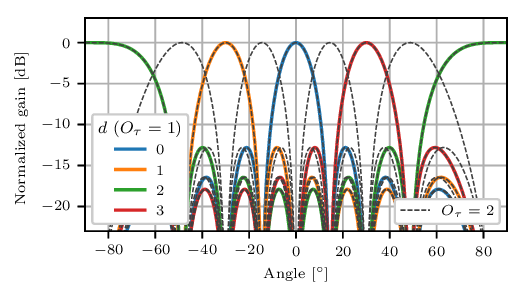}
    \caption{Beampatern of the main harmonic component $(i=0)$ per delay value for $M=8$ antennas, $N=4$ and two values of pulse duration scaling factor $O_{\mathrm{\tau}} = 1, 2$. }
    \label{fig:beampat_per_delay_oversamp}
\end{figure}

The considered ULA TMA array operates in the far field with isotropic  antenna elements spaced by $d_{\mathrm{a}}$. Taking the first element as a reference the path difference per antenna element in the direction $\theta$ is
\begin{align}
    \label{eq:path_diff_ped_ant}
    \Delta d_m &= m d_{\mathrm{a}} \sin{(\theta)} \nonumber \\
    &= m d_{\lambda} \frac{c}{f_{\mathrm{c}}} \sin{(\theta)},
\end{align}
where $d_{\lambda}$ is the spacing between antenna elements expressed in wavelengths and $f_{\mathrm{c}}$ is the center frequency that the array is designed at.
To achieve beamforming, the modulating sequence per antenna is cyclically shifted by $m d$, resulting in the phase difference between adjacent antennas equal to $2 \pi d / D$. The phase shift at the $m$-th antenna according to \eqref{eq:ph_shift_per_harm_per_delay} is 
\begin{align}
    \label{eq:phase_shift_per_antenna}
    \phi_m(i,d) 
                &=-2\pi m d \left( \frac{1}{D} + \frac{i}{O_{\mathrm{\tau}}} \right).
\end{align}
The array factor (AF) of the TMA at the $i$-th harmonic frequency for a selected delay $d$ is 
\begin{align}
    \label{eq:tma_array_factor}
    AF(\theta, i, d) &= \alpha(i) \frac{1}{\sqrt{M}}  \sum_{m=0}^{M-1} e^{j \phi_m(i, d)} e^{-j 2\pi  \frac{\Delta d_m}{c} \left(f_{\mathrm{c}} + \frac{f_{\mathrm{p}}}{N} + i f_{\mathrm{p}} \right) },
\end{align}
where $1/\sqrt{M}$ is the power normalization factor.
Assuming that the frequencies of the harmonic components with significant power are negligible compared to the carrier frequency $\left( \frac{1}{N} + i\right) f_{\mathrm{p}} \ll f_{\mathrm{c}}$ the AF can be simplified to
\begin{align}
    \label{eq:tma_af_simplified}
    AF(\theta, i, d) 
    &= \alpha(i) \frac{1}{\sqrt{M}} \sum_{m=0}^{M-1} e^{-2\pi m \left( \frac{d}{D} \left( 1 + Ni\right) + d_{\lambda} \sin{(\theta)} \right) }
\end{align}
The TMA achieves analog (constant) precoding over each harmonic replica of the baseband signal by cyclically shifting the modulating sequence. Depending on the harmonic index and delay, the formed beams are pointed in directions $\theta(i, d) = -\arcsin{ \left( \frac{d}{D d_{\lambda}} \left(  1 +iN \right) \right)},\ i\in \mathbb{Z}$. As a result, the harmonic beams might be pointed in different directions than the main harmonic $(i=0)$ beam. Fig. \ref{fig:beampat_per_delay_oversamp} shows the beam pattern of the main harmonic $(i=0)$ per selected delay value $d$ for two pulse duration extension factors. The oversampling factor $O_{\mathrm{\tau}}$ effectively increases the phase shift resolution, improving the accuracy of the beamforming.

\subsection{Array tapering}

\begin{figure}[b]
    \centering
    \includegraphics[width=3.5in]{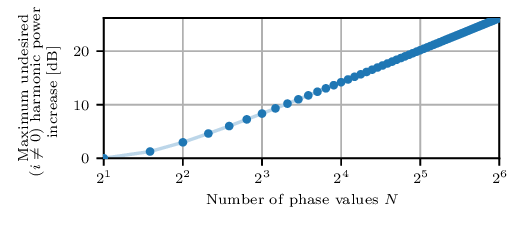}
    \caption{Worst-case undesired harmonic $(i \neq 0)$ power increase while tapering as compared to the case without tapering as a function of the number of phase shift values $N$.}
    \label{fig:h_gain_vs_n_phase_vals}
\end{figure}

The oversampled modulating signal can be considered for array tapering provided that the RF front-end switch has an additional off-state. The tapering is achieved by introducing zeros in the modulating pulse.
This allows to reduce the power of the transmitted signal that can be adjusted per antenna element facilitating spatial windowing. Given the pulse duration scaling factor, the number of samples per pulse is $O_{\mathrm{\tau}}$, which allows us to obtain tapering amplitudes $O_{\mathrm{\tau}} + 1$ that are uniformly distributed from $0$ to $1$. Fig. \ref{fig:n_bits_vs_oversamp} shows the number of bits of the tapering amplitude as a function of the oversampling factor $O_{\mathrm{\tau}}$.
The pulse duration with tapering is
\begin{equation}
        \label{eq:tap_pulse_dur}
    T_{\mathrm{pt}} = T_{\mathrm{p}} \frac{O_{\mathrm{\tau}} - l}{O_{\mathrm{\tau}}} = T_{\mathrm{p}} \eta(l),
\end{equation}
where $l$ is the number of zeros and $\eta(l)$ is the pulse duration scaling coefficient due to tapering.
By revisiting \eqref{eq:fourier_coeffs_pulse_seq} the harmonic components of the tapered modulating sequence are
\begin{align}
    \label{eq:fourier_coeffs_w_tapering}
    C(k) &= \eta(l) \sinc{\left(\pi \frac{k}{N} \eta(l) \right)} e^{-j\pi \frac{k}{N} \eta(l)} I(k).
\end{align}
Due to the shortened pulse duration, the tapering affects the power and phase of the harmonic components. When considering beamforming with tapering, the additional phase shift must be accounted for in the design of the delays per antenna. Moreover, some tapering values might result in the increased power of the undesired harmonic components $(i \neq 0)$ as compared to the case without it, due to shortened pulse duration. Based on numerical simulations, the increase in harmonic power depends on the number of phase shift values of the modulating sequence $N$. Fig. \ref{fig:h_gain_vs_n_phase_vals} presents the maximum (worst-case) increase in the power of the harmonic components across all tapering amplitudes as a function of the number of phase shift values. Increasing the number of phase shifts twofold results in the 6dB worst-case power gain of the undesired harmonic components.

\section{Conclusion}
This paper analyzes the TMA with constrained switching frequency and quantized modulating sequence. The power and spacing of the harmonic components generated by the modulation are investigated in detail. The presented signal model illustrates the trade-offs offered by different use of the oversampling factor when synthesizing the modulating signal. By extending the pulse duration by the factor $O_{\mathrm{\tau}}$, the effective phase shifting resolution is improved $O_{\mathrm{\tau}}$ times. Furthermore, the oversampled pulse allows one to consider tapering by adjusting the power of the modulating signal, facilitating spatial windowing.

\bibliographystyle{IEEEtran}
\bibliography{biblio.bib}

\begin{thebibliography}{10}
\providecommand{\url}[1]{#1}
\csname url@samestyle\endcsname
\providecommand{\newblock}{\relax}
\providecommand{\bibinfo}[2]{#2}
\providecommand{\BIBentrySTDinterwordspacing}{\spaceskip=0pt\relax}
\providecommand{\BIBentryALTinterwordstretchfactor}{4}
\providecommand{\BIBentryALTinterwordspacing}{\spaceskip=\fontdimen2\font plus
\BIBentryALTinterwordstretchfactor\fontdimen3\font minus \fontdimen4\font\relax}
\providecommand{\BIBforeignlanguage}[2]{{%
\expandafter\ifx\csname l@#1\endcsname\relax
\typeout{** WARNING: IEEEtran.bst: No hyphenation pattern has been}%
\typeout{** loaded for the language `#1'. Using the pattern for}%
\typeout{** the default language instead.}%
\else
\language=\csname l@#1\endcsname
\fi
#2}}
\providecommand{\BIBdecl}{\relax}
\BIBdecl

\bibitem{6g_vision}
H.~Tataria, M.~Shafi, A.~F. Molisch, M.~Dohler, H.~Sjöland, and F.~Tufvesson, ``{6G} wireless systems: Vision, requirements, challenges, insights, and opportunities,'' \emph{Proceedings of the IEEE}, vol. 109, no.~7, pp. 1166--1199, 2021.

\bibitem{unconvent_array_arch}
P.~Rocca, G.~Oliveri, R.~J. Mailloux, and A.~Massa, ``Unconventional phased array architectures and design methodologies—a review,'' \emph{Proceedings of the IEEE}, vol. 104, no.~3, pp. 544--560, 2016.

\bibitem{rfic_book}
F.~Ellinger, \emph{Radio Frequency Integrated Circuits and Technologies}.\hskip 1em plus 0.5em minus 0.4em\relax Springer Publishing Company, Incorporated, 01 2007.

\bibitem{mmw_tech_book}
A.~M. Niknejad and H.~Hashemi, \emph{Mm-Wave Silicon Technology: 60 {GHz} and Beyond}, 1st~ed.\hskip 1em plus 0.5em minus 0.4em\relax Springer Publishing Company, Incorporated, 2008.

\bibitem{overview_tma_fda}
W.-Q. Wang, H.~C. So, and A.~Farina, ``An overview on time/frequency modulated array processing,'' \emph{IEEE Journal of Selected Topics in Signal Processing}, vol.~11, no.~2, pp. 228--246, 2017.

\bibitem{tma_survey}
L.~P. P.~Rocca, F.~Yang and S.~Yang, ``Time-modulated array antennas – theory, techniques, and applications,'' \emph{Journal of Electromagnetic Waves and Applications}, vol.~33, no.~12, pp. 1503--1531, 2019.

\bibitem{tma_beamscanning_init}
H.~Shanks, ``A new technique for electronic scanning,'' \emph{IRE Transactions on Antennas and Propagation}, vol.~9, no.~2, pp. 162--166, 1961.

\bibitem{tma_seq_sr_optimization}
S.~Yang, Y.~B. Gan, A.~Qing, and P.~K. Tan, ``Design of a uniform amplitude time modulated linear array with optimized time sequences,'' \emph{IEEE Transactions on Antennas and Propagation}, vol.~53, no.~7, pp. 2337--2339, 2005.

\bibitem{phase_tma_init}
J.~Yang, W.~Li, and X.~Shi, ``Phase modulation technique for four-dimensional arrays,'' \emph{IEEE Antennas and Wireless Propagation Letters}, vol.~13, pp. 1393--1396, 2014.

\bibitem{single_sideband_phase_iq_tma}
A.-M. Yao, W.~Wu, and D.-G. Fang, ``Single-sideband time-modulated phased array,'' \emph{IEEE Transactions on Antennas and Propagation}, vol.~63, no.~5, pp. 1957--1968, 2015.

\bibitem{tma_ampl_phase_weighting_beamfomring}
H.~Li, Y.~Chen, and S.~Yang, ``Harmonic beamforming in antenna array with time-modulated amplitude-phase weighting technique,'' \emph{IEEE Transactions on Antennas and Propagation}, vol.~67, no.~10, pp. 6461--6472, 2019.

\bibitem{lin_phase_approx_tma_eff_analysis}
Q.~Zeng, P.~Yang, L.~Yin, H.~Lin, C.~Wu, F.~Yang, and S.~Yang, ``Phase modulation technique for harmonic beamforming in time-modulated arrays,'' \emph{IEEE Transactions on Antennas and Propagation}, vol.~70, no.~3, pp. 1976--1988, 2022.

\bibitem{high_res_phase_shifting_w_tma}
Y.~Gao, X.~Liang, J.~Chen, J.~Chen, and R.~Jin, ``A high-resolution amplitude-phase control method for 2-bit time-modulated phased arrays,'' \emph{IEEE Transactions on Antennas and Propagation}, vol.~71, no.~11, pp. 8692--8703, 2023.

\bibitem{tma_pwr_freq_analysis}
J.~C. Bregains, J.~Fondevila-Gomez, G.~Franceschetti, and F.~Ares, ``Signal radiation and power losses of time-modulated arrays,'' \emph{IEEE Transactions on Antennas and Propagation}, vol.~56, no.~6, pp. 1799--1804, 2008.

\end{thebibliography}

\end{document}